\documentclass{aa}
\usepackage[varg]{txfonts}
\usepackage{color}
\usepackage{graphicx}
\usepackage{natbib}
\usepackage{epstopdf}
\bibpunct{(}{)}{;}{a}{}{,} 

\begin{document}

\title{Seismological constraints on the solar coronal heating function}
\author{D.~Y.~Kolotkov\inst{1,2}\thanks{Corresponding~author: D.Kolotkov.1@warwick.ac.uk} \and T.~J.~Duckenfield\inst{1}\and V.~M.~Nakariakov\inst{1,3}}

\authorrunning{Kolotkov et al.}
\titlerunning{Constraining the solar coronal heating function}

\institute{Centre for Fusion, Space and Astrophysics, Department of Physics, University of Warwick, CV4 7AL, UK\label{1}
\and
Institute of Solar-Terrestrial Physics SB RAS, Irkutsk 664033, Russia \label{2}
\and
St. Petersburg Branch, Special Astrophysical Observatory, Russian Academy of Sciences, 196140, St. Petersburg, Russia\label{3}
}

\date{Received \today /Accepted dd mm yyyy}

\abstract
{}
{The hot solar corona exists because of the balance between radiative and conductive cooling and some counteracting heating mechanism which remains one of the major puzzles in solar physics.}
{The coronal thermal equilibrium is perturbed by magnetoacoustic waves which are abundantly present in the corona, causing a misbalance between the heating and cooling rates. Due to this misbalance, the wave experiences a back-reaction, either losing or gaining energy from the energy supply that heats the plasma, at the time scales comparable to the wave period.}
{In particular, the plasma can be subject to wave-induced instability or over-stability, depending on the specific choice of the coronal heating function. In the unstable case, the coronal thermal equilibrium would be violently destroyed, which does not allow for the existence of long-lived plasma structures typical for the corona. Based on this, we constrained the coronal heating function using observations of slow magnetoacoustic waves in various coronal plasma structures.}
{}


\maketitle

\section{Introduction}

The existence of the one million Kelvin solar corona requires a continuous supply of energy to compensate energy losses by optically thin radiation and field-aligned thermal conduction down to the chromosphere \citep{2015RSPTA.37340269D}. Otherwise, i.e. without re-supply of energy by some heating mechanism, due to radiative losses acting alone the corona would cool down on the time scale
\begin{equation}\label{eq:tau_rad}
\tau_\mathrm{rad} = \frac{\gamma C_\mathrm{V}T_0}{{\cal L}_0(\rho_0, T_0)}.
\end{equation}
Here, $\gamma$ is the adiabatic index, $C_\mathrm{V}$ is the specific heat capacity {see Table~\ref{tab:params}}, and ${\cal L}_0$ is the radiative loss function depending on temperature, $T_0$ and density, $\rho_0$ and measured in W\,kg$^{-1}$ in this work. Estimations show that for typical coronal conditions the radiative cooling time scale $\tau_\mathrm{rad}$ (\ref{eq:tau_rad}) ranges approximately as $10^3$--$10^4$\,s \citep{2004A&A...415..705D,2008ApJ...686L.127A,2018AdSpR..61..645P}, that is substantially shorter than the observed lifetime of typical coronal structures.  Revealing the nature of such a mechanism for retaining the hot coronal temperature constitutes the long-standing \textit{coronal heating problem}. In this work, we determine empirical constraints on the coronal heating function by the observed behaviour of slow magnetoacoustic waves. 

There is abundant evidence of the existence of slow magnetoacoustic waves in the corona \citep{2009SSRv..149...65D,2011SSRv..158..267B,2011SSRv..158..397W,2012RSPTA.370.3193D}, which almost always show rapid damping. In particular, rapidly damped slow waves are observed in hot and warm coronal loops \citep{2011SSRv..158..397W,2016GMS...216..395W,2019ApJ...874L...1N}, and coronal holes \citep{2011SSRv..158..267B,2016GMS...216..419B}. Thus, thermodynamical properties of those plasma structures, including heating, must lead to the wave behaviour consistent with the observations. Coronal slow waves are sensitive to effects of thermal conduction, optically thin radiation, and compressive viscosity \citep{2003A&A...408..755D,2004A&A...415..705D,2005A&A...436..701S,2014ApJ...789..118K,2016ApJ...820...13M,2018MNRAS.478..342B}. Historically, thermal conduction is invoked as the dominant mechanism for the rapid damping of coronal slow waves \citep{2002ApJ...580L..85O,2003A&A...408..755D}. Yet observations measuring a change in the effective polytropic index with temperature show that in some cases thermal conduction alone is insufficient to describe the damping \citep{2009A&A...494..339O,2015ApJ...811L..13W,2018ApJ...868..149K}. The measured phase shifts between density and temperature perturbations in a slow wave are different to the phase shift predicted to arise from thermal conduction in \citet{2009A&A...494..339O}. 
This discrepancy is supported by the apparent lack of the decrease in damping length with temperature, as would be expected for thermal conduction \citep{2019FrASS...6...57S}. Combined with different dependencies of the damping time upon wave period, observed in warm coronal loops and cooler plumes \citep{2014ApJ...789..118K} and at different heights \citep{2014A&A...568A..96G}, this demonstrates the need for accounting for an additional damping mechanism. 

For compressive waves, the coronal plasma that appears to be in thermal equilibrium formed by the balance of energy losses and gains, acts as an \emph{active medium}. That is to say, the wave can gain energy from the medium. In other words, the wave is subject to a back-reaction of the modification of the cooling and heating rates by the density and temperature perturbations in the wave, i.e., a wave-induced \emph{heating/cooling misbalance}. This effect can cause enhanced damping or amplification of slow waves \citep{2000ApJ...528..767N,2006A&A...460..573C,2016ApJ...824....8K,2017ApJ...849...62N,2018PhPl...25d2116P,2019A&A...624A..96C,2019A&A...628A.133K}. Characteristic time scales of the thermal misbalance lead to the wave dispersion \citep{1993ApJ...415..335I,2019PhPl...26h2113Z}. The specific behaviour depends on the equilibrium parameters and derivatives of the combined heating/cooling function with respect to density and temperature (in the linear regime). In addition to a slow wave, there exists also a non-propagating entropy mode. This mode causes thermal misbalance and is affected by the back-reaction too, leading, in particular, to radiative instability \citep{1965ApJ...142..531F}. When physical conditions for such a thermal instability are fulfilled, rapid condensations of the coronal plasma and the phenomenon of coronal rain may occur \citep{1988SoPh..117...51D,2010ApJ...716..154A}.

The importance of the plasma activity depends on the ratio of the characteristic times of thermal misbalance and wave oscillation periods. This question has so far received only limited attention \citep[cf. the previous works by][on thermal misbalance]{2016ApJ...824....8K,2017ApJ...849...62N,2019A&A...628A.133K,2019PhPl...26h2113Z}.
Here we demonstrate that the characteristic time scales of the thermal misbalance are about typical periods of slow magnetoacoustic waves observed in various coronal plasma structures, such as polar plumes and quiescent and flaring loops. As such, the dynamics of slow waves is proved to be highly sensitive to the discussed thermodynamic activity of the corona, including thermodynamic properties of the enigmatic heating mechanisms. We perform quantitative assessment of the acoustic and thermal instability criteria in a broad range of coronal conditions and for various heating models. We identify heating models in which the plasma is either stable to thermal instability and slow magnetoacoustic over-stability, or the growth rate is relatively low, allowing for the existence of long-lived plasma structures in the corona and causing {the} rapid damping of slow magnetoacoustic waves, {observed in the corona}.

\section{Governing equations and dispersion relation}
In the infinite magnetic field approximation with plasma $\beta\to 0$, slow magnetoacoustic waves propagate strictly along the field and their dynamics in a heated plasma is governed by the set of one-dimensional equations,
\begin{align}\label{eq_motion}
	&\rho \frac{d V_{z}}{d t} = -\frac{\partial P}{\partial z},\\
	&\frac{\partial \rho}{\partial t} +  \frac{\partial }{\partial z} \left(\rho V_{z} \right) = 0,\\
	&P=\frac{k_\mathrm{B} T \rho}{m},\\
	&C_\mathrm{V}\frac{dT}{dt} - \frac{k_\mathrm{B}T}{m\rho}\frac{d\rho}{dt}=-{\cal Q}(\rho,T)+\frac{\kappa}{\rho}\frac{\partial^2T}{\partial z^2},\label{eq:energy_full}
\end{align}
where $\rho$, $T$, and $P$ are the density, temperature, and pressure, respectively; $ V_{z}$ is the velocity along the \textit{z}-axis directed along the magnetic field. Equilibrium parameters and constant quantities are given in Table~\ref{tab:params}{, and $d/dt$ stands for the total derivative}.
{Below, we apply this model to both standing and propagating slow waves in coronal loops. For standing SUMER-type oscillations in hot flaring loops, the stratification scale height is large enough allowing us to neglect gravitational stratification \citep[e.g.][]{2018ApJ...860..107W, 2019ApJ...886....2W}. In the quiescent one million Kelvin loops though, the loop height is comparable to the gravitational scale height, but the propagating slow waves damp near the very bottom of the loop \citep[e.g.][]{2006A&A...448..763M} where the effects of stratification are not yet fully developed. We also neglect the effect of viscosity as in terms of regulating the coronal thermal equilibrium, thermal conduction is of greater importance than viscosity \citep[see e.g.][]{2003A&A...408..755D}, and so we choose to include thermal conduction with the expectation that viscosity plays less influential role. Also, to keep the application of this seismological approach as simple and generic as possible, the effects of anomalous plasma transport coefficients \citep[e.g.][]{2015ApJ...811L..13W} and plasma flows are not considered in this work.}

Terms on the right-hand side of energy equation (\ref{eq:energy_full}) represent non-adiabatic processes which are some unspecified heating ${\cal H}(\rho,T)$ and optically thin radiative cooling ${\cal L}(\rho,T)$, combined in the net heat/loss function ${\cal Q}(\rho,T) = {\cal L} - {\cal H}$; and the field-aligned thermal conduction with the coefficient $\kappa$. In the equilibrium, ${\cal Q}_0=0$ and the plasma temperature is uniform. 
{The assumption of the isothermal equilibrium along the coronal part of the loop is, on one hand, motivated by observations \citep[e.g.][]{2015ApJ...800..140G, 2016ApJ...828...72M, 2017A&A...600A..37N, 2019A&A...627A..62G}, and, on the other hand, consistent with a common approach in theoretical modelling of coronal loop oscillations \citep[e.g.][]{2005A&A...436..701S, 2009A&A...494..339O, 2016ApJ...820...13M, 2018ApJ...860..107W, 2019ApJ...886....2W}. Hence, in the equilibrium, the second term on the right-hand side of Eq.~(\ref{eq:energy_full}) is zero.}
Thus, in addition to the perturbation of the mechanical equilibrium provided by the force balance, we consider a perturbation of the thermal equilibrium, i.e., allow both heating and cooling processes to be perturbed by the wave. The latter causes an additional mechanism for the energy exchange between the plasma and the wave, that is the {heating/cooling misbalance}.

In the linear regime and assuming the harmonic dependence of the wave-perturbed plasma parameters upon time and spatial coordinate in Eqs.~(\ref{eq_motion})--(\ref{eq:energy_full}), the relation between the cyclic frequency $\omega$ and wavenumber $k$, written in terms of the characteristic time scales of the considered non-adiabatic processes, is
\begin{equation}\label{eq:dr}
\omega^3 + iA(k)\omega^2 - B(k)\omega - iC(k) = 0,
\end{equation}
with $A={\tau_{\mathrm{cond}}^{-1}}+{\tau_2^{-1}}$, $B=C_\mathrm{s}^2k^2$, $C = \left(\tau_{\mathrm{cond}}^{-1}+{\gamma}{\tau_1^{-1}}\right) C_\mathrm{s}^2k^2/\gamma$,
where $C_\mathrm{s}=\sqrt{\gamma k_\mathrm{B}T_0/m}$ is the sound speed and $\tau_{\mathrm{cond}}={\rho_0 C_\mathrm{V}k^{-2}}/{\kappa}$
is the characteristic time of the wavelength-dependent thermal conduction. Likewise, $\tau_{1}={\gamma C_\mathrm{V}}/\left[{{\cal Q}_{T}-(\rho_0/T_0){\cal Q}_{\rho}}\right]$ and $\tau_{2}={C_\mathrm{V}}/{{\cal Q}_{T}}$ in Eq.~(\ref{eq:dr}) are the time scales describing rates of change of the perturbed heat/loss function ${\cal Q}$ with plasma density ${\cal Q}_\mathrm{\rho}\equiv(\partial {\cal Q}/\partial \rho)_T $ and temperature ${\cal Q}_{T} \equiv \left( \partial {\cal Q}  / \partial T \right)_{\rho}$, respectively.
{This dispersion relation was derived in \citet{2019A&A...628A.133K} and \citet{2019PhPl...26h2113Z}.}
Being the third-order polynomial in $\omega$, dispersion relation (\ref{eq:dr}) describes two slow waves coupled with a thermal (entropy-related) mode.

\begin{table}
	\centering
	\caption{Typical parameters of the coronal plasma and slow waves, used for the analysis.
		The wave periods $\tau$ are shown for propagating waves observed in loops \citep{2009SSRv..149...65D} with $\tau \simeq 5$\,min and plumes \citep{2011SSRv..158..267B} with $\tau \simeq 15$\,min, and standing SUMER oscillations  \citep{2011SSRv..158..397W,2019ApJ...874L...1N} with $\tau \simeq 15$\,min. {The constant $k_\mathrm{B}$ is the Boltzmann constant.}
		\label{tab:params}
	}
	\begin{tabular}{ll}
		\\
		\hline
		Parameter & Value \\
		\hline
		Temperature, $T_0$ & 0.5--20\,MK  \\
		Number density, $n_0$ & $10^8$--$10^{11}$\,cm$^{-3}$\\
		Wave periods, $\tau$ & 5 min and 15 min \\
		Spitzer conductivity, $\kappa$ & $10^{-11}T_0^{5/2}\,\mathrm{W\,m}^{-1}\,\mathrm{K}^{-1}$\\
		Adiabatic index, $\gamma$ & 5/3\\
		Mean particle mass, $m$ & $0.6\times1.67\times10^{-27}\,\mathrm{kg}$\\
		{Specific heat capacity,} $C_\mathrm{V}$ & $(\gamma-1)^{-1}k_\mathrm{B}/m$ $\mathrm{J}\,\mathrm{K}^{-1}\,\mathrm{kg}^{-1}$\\
		\hline
	\end{tabular}
\end{table}

\section{Stability of acoustic and thermal modes for varying heating models}

In the infinite magnetic field approximation, slow and thermal modes become unstable and gain energy from the plasma if \citep{1965ApJ...142..531F,2019PhPl...26h2113Z,2019A&A...628A.133K}
\begin{align}\label{eq:instab_cond_ac}
	&\frac{1}{\tau_2}-\frac{1}{\tau_1}+\frac{\gamma-1}{\gamma}\frac{1}{\tau_\mathrm{cond}}<0, &\mathrm{for~acoustic~mode}\\
	&\frac{\gamma}{\tau_1}+\frac{1}{\tau_\mathrm{cond}}<0, &\mathrm{for~thermal~mode}\label{eq:instab_cond_th}
\end{align}
where $\tau_\mathrm{cond}$, $\tau_1$, and $\tau_2$ are the characteristic wavelength-dependent time of the field-aligned thermal conduction, and wavelength-independent time scales of the heating/cooling misbalance, determined in Eq.~(\ref{eq:dr}).
The thermal conduction time $\tau_\mathrm{cond}$ is essentially positive and therefore contributes always to damping. In contrast, the values of $\tau_{1,2}$, as well as their difference, can be either positive or negative depending on the equilibrium plasma parameters and properties of the heating and cooling functions. 
{In general, negative characteristic times correspond to the perturbation growing in time. We discriminate between the overstable regime, which refers to an oscillation with a growing amplitude, and the unstable regime, in which the perturbation grows aperiodically.}
Table~\ref{tab:regimes} summarises qualitative behaviours of the thermal and acoustic modes {for {all} six possible combinations of positive and negative $\tau_{1,2}$ and $\tau_1 - \tau_2$.}

\begin{table*}
	\centering
	\caption{Possible regimes of the thermal and acoustic modes in the corona, determined by Eqs.~(\ref{eq:instab_cond_ac})--(\ref{eq:instab_cond_th}).
		The weaker/stronger damping is defined in comparison with that caused by thermal conduction alone.
		\label{tab:regimes}
	}
	\begin{tabular}{ccccc}
		\\
		\hline
		$\tau_1$ & $\tau_2$ & $\tau_1-\tau_2$& Thermal mode & Acoustic mode\\
		\hline
		$>0$ & {$>0$} &$>0$& damps stronger & damps stronger \\[-0.1cm]
		&  &$<0$&damps stronger  & damps weaker/overstable \\
		$<0$ & {$<0$} &$>0$& damps weaker/unstable  & damps stronger \\[-0.1cm]
		&  &$<0$& damps weaker/unstable & damps weaker/overstable \\
		$>0$ & $<0$ & $>0$&damps stronger & damps weaker/overstable \\
		$<0$ & $>0$ &$<0$& damps weaker/unstable & damps stronger \\
		\hline
	\end{tabular}
\end{table*}

We synthesise the coronal optically thin radiation function ${\cal L}(\rho,T)$ from CHIANTI atomic database v.~9.0.1 \citep{1997A&AS..125..149D, 2019ApJS..241...22D} for the ranges of densities and temperatures shown in Table~\ref{tab:params}.
The coronal heating function is parametrised as ${\cal H}(\rho,T)=h\rho^aT^b$ where the coefficient $h$ is determined from the initial thermal equilibrium condition, ${\cal Q}_0=0$, as $h={\cal L}_0/\rho_0^aT_0^b$; and the power indices $a$ and $b$ are treated as free parameters.
{In this parametrisation, the dependence of the heating function on the magnetic field strength is omitted, as it was shown to have no effect on the slow wave dynamics in the zero-$\beta$ limit considered in our work \citep[see Eqs.~(27) and (28) in][]{2017ApJ...849...62N}.}
Similar forms of the heating model were considered in, e.g., \citet{1988SoPh..117...51D,1993ApJ...415..335I,2000ApJ...530..999M,2006A&A...460..573C}. This allows us to re-write conditions (\ref{eq:instab_cond_ac})--(\ref{eq:instab_cond_th}) in terms of the heating power indices $a$ and $b$ as
\begin{align}\label{eq:instab_cond_ac_ab}
	&\mathrm{(for~acoustic~overstability)}\\
	&b > -\frac{a}{\gamma-1} + \frac{T_0}{{\cal L}_0}\left(\frac{\partial {\cal L}}{\partial T}\right)_{T_0,\rho_0} + \frac{C_\mathrm{V}}{\tau_\mathrm{cond}}\frac{T_0}{{\cal L}_0} +\frac{1}{\gamma-1},&\nonumber\\
	&\mathrm{(for~thermal~instability)}\label{eq:instab_cond_th_ab}\\
	&b > a + \frac{T_0}{{\cal L}_0}\left(\frac{\partial {\cal L}}{\partial T}\right)_{T_0,\rho_0} + \frac{C_\mathrm{V}}{\tau_\mathrm{cond}}\frac{T_0}{{\cal L}_0} - 1.&\nonumber
\end{align}
These conditions allow us to delineate different coronal heating models, i.e., values of $a$ and $b$, for which slow magnetoacoustic and thermal modes either damp or grow, that could be directly verified in observations.

\begin{figure}
	\centering
	\includegraphics[width=0.392\linewidth]{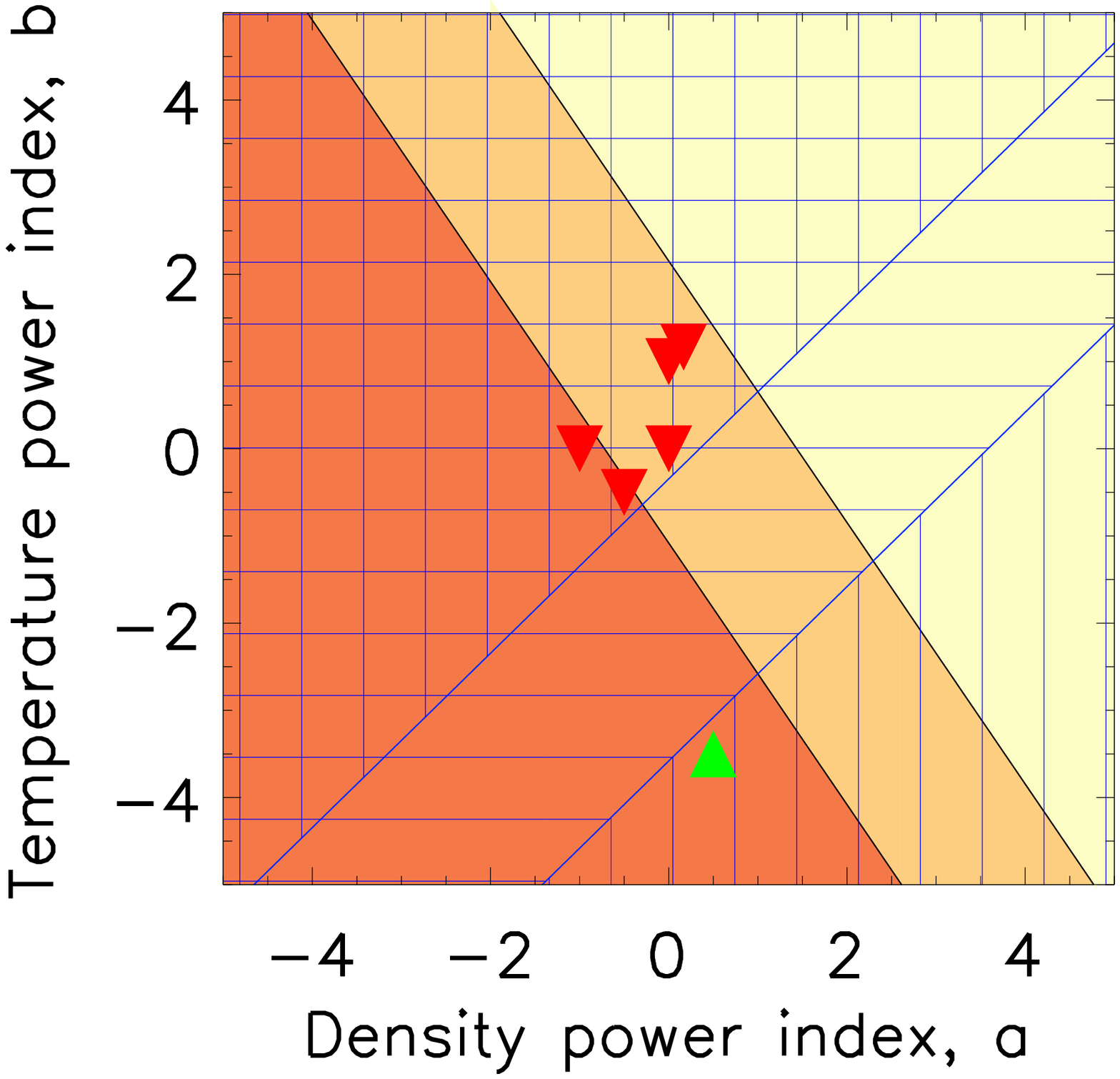}
	\includegraphics[width=0.392\linewidth]{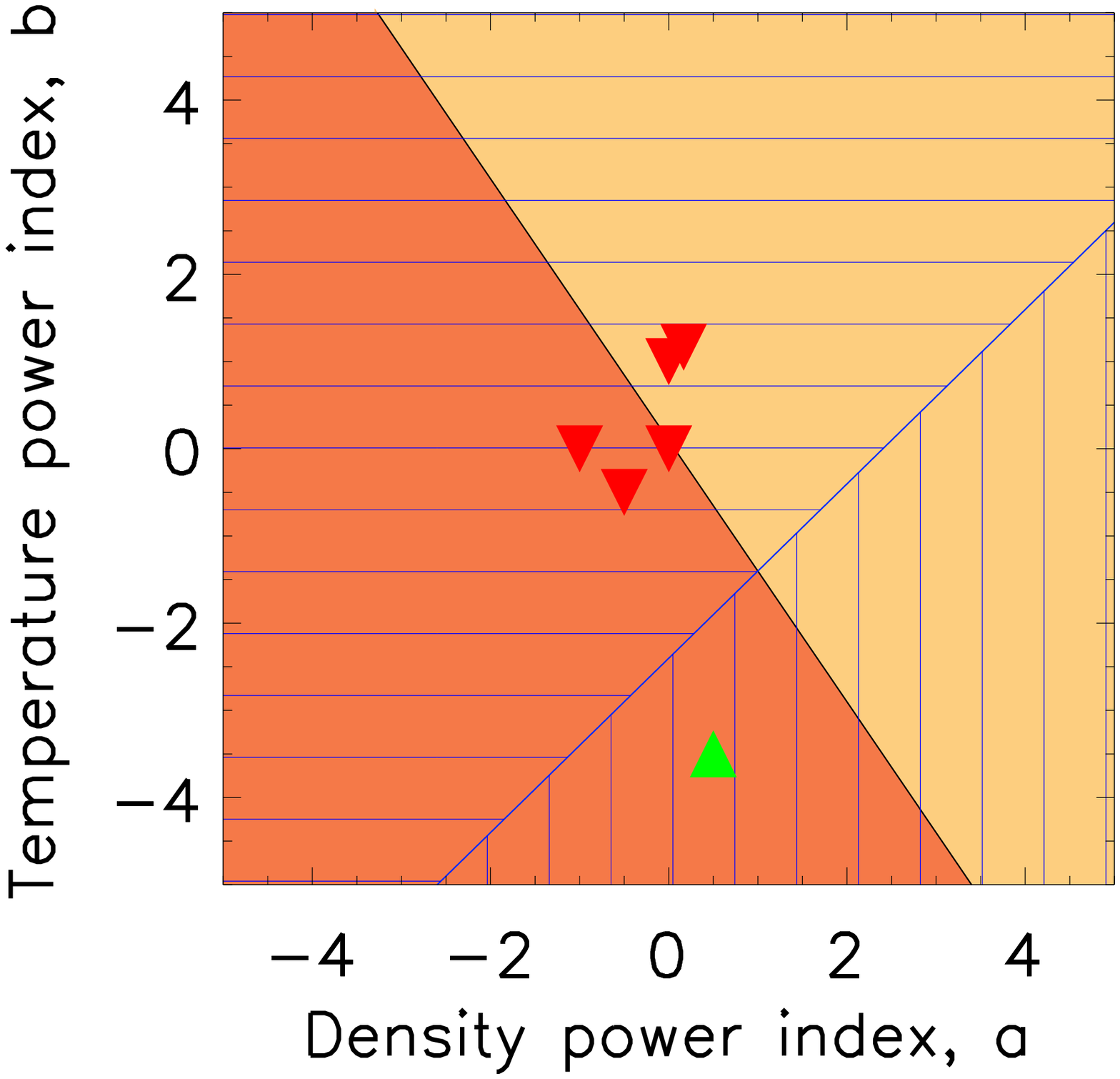}
	\includegraphics[width=0.197\linewidth]{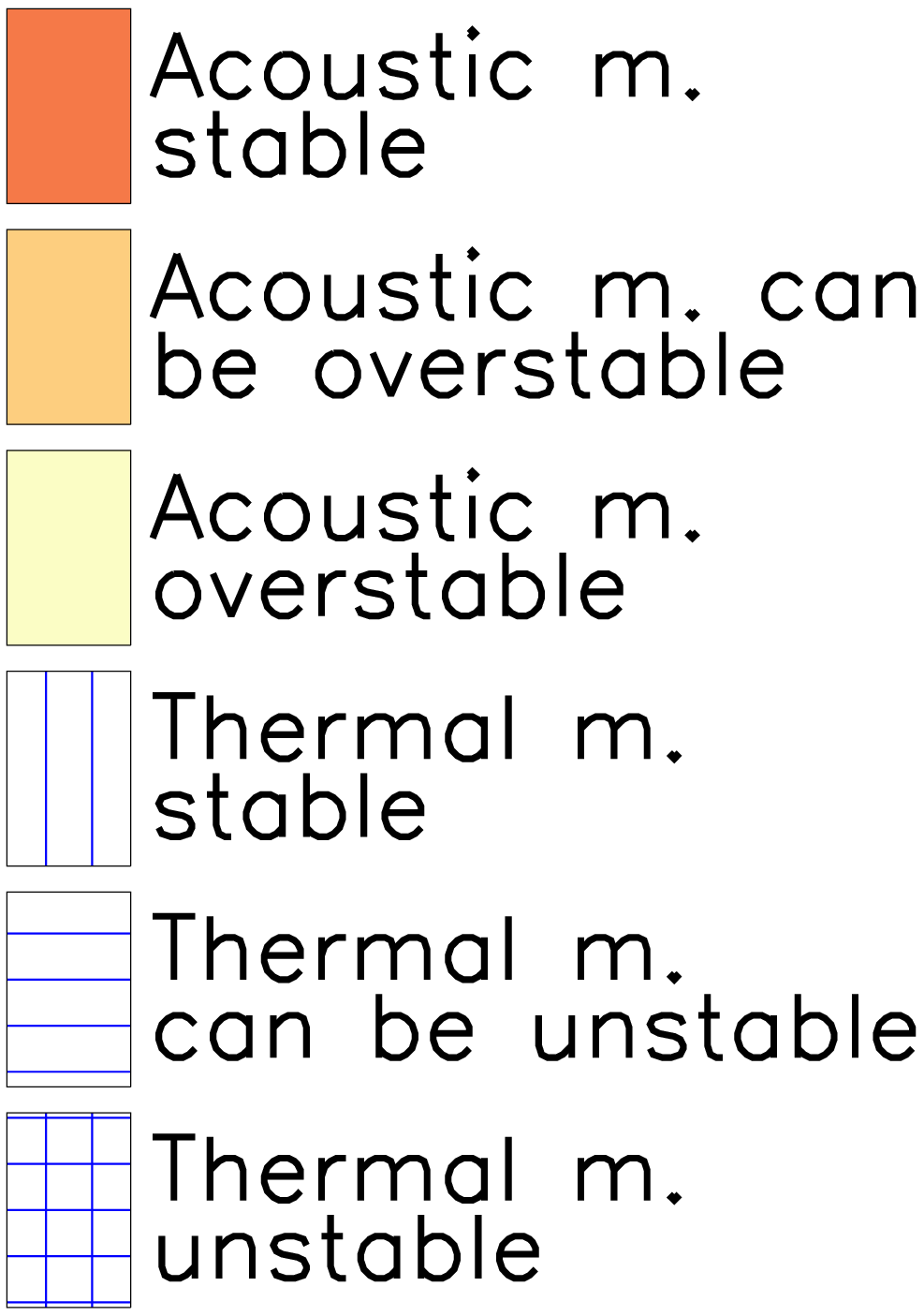}
	\caption{Power indices $a$ and $b$ in the parametrisation of the coronal heating function as ${\cal H}(\rho,T)\propto \rho^aT^b$, determining stability of acoustic and thermal modes with (right panel) and without (left) thermal conduction. The red triangles show the values of $a$ and $b$ for the heating models from \citet{1993ApJ...415..335I}. {Note the difference in the units of the heating function used in this work $\mathrm{[W\,kg}^{-1}$] \citep[see also][]{1965ApJ...142..531F} and in \citet{1993ApJ...415..335I} $[\mathrm{W\,m}^{-3}]$, leading to the difference in the density index $a$ by 1.} The green triangle shows the heating model used as an example in Fig.~\ref{fig:tau_rho_T}.
	}
	\label{fig:ab_regions}
\end{figure}

First, we analyse Eqs.~(\ref{eq:instab_cond_ac_ab})--(\ref{eq:instab_cond_th_ab}) in the regime of suppressed thermal conduction, taking $\tau_\mathrm{cond}\to \infty$ \citep[see e.g.][]{2015ApJ...811L..13W,2018ApJ...860..107W}. 
We obtain instability thresholds for a broad range of coronal plasma densities and temperatures (Table \ref{tab:params}) typical for hot jets, plumes, and interplume regions in coronal holes \citep{2009LRSP....6....3C,2011A&ARv..19...35W}; warm quiescent coronal loops and hot and dense flaring loops \citep{2014LRSP...11....4R}. 
The thresholds are manifested as two intersecting families of parallel straight lines with the slopes of $-{(\gamma-1)^{-1}}$ and $1$ in the parametric plane $(a,b)$. From both sets, we pick the outermost boundaries that allow us to identify intervals of $a$ and $b$ for which thermal and acoustic modes appear in different regimes (Fig.~\ref{fig:ab_regions}, left panel).

In the case with a finite Spitzer thermal conductivity, for a characteristic {wavelength of the perturbation $\lambda \simeq 360$\,Mm \citep[typical of standing slow waves,][]{2011SSRv..158..397W}}, those critical intervals get softened (Fig.~\ref{fig:ab_regions}, right panel), and the regions of omnipresent thermal and acoustic instabilities disappear. The increase in $\lambda$ weakens the effect of thermal conductivity.  However, for all plausible values of $\lambda$ extending up to 2000\,Mm, no change to the behaviour of the acoustic and thermal modes shown in the right-hand panel in Fig.~\ref{fig:ab_regions} was detected.
Likewise, the effect of the thermal conductivity strengthens for shorter {$\lambda$ (typical of propagating slow waves in quiescent loops, for example)}, pushing the regions of omnipresent instabilities further away from the considered intervals of $a$ and $b$. Thus, both modes are found to damp through the whole considered intervals of plasma densities and temperatures for the values of $a$ and $b$ in the region outlined by a triangle with approximate vertices $(-2.5,-5)$, $(1,-1.5)$, and $(3.5,-5)$. In other words, for all heating models with $a$ and $b$ in this region the slow mode will be seen in observations as a damped wave in a thermally stable plasma. For example, for the heating models suggested by \citet{1978ApJ...220..643R,1993ApJ...415..335I},  including Ohmic heating ($a=0$, $b=1$), constant heating per unit volume ($a=-1$, $b=0$) and mass ($a=0$, $b=0$), and heating by Alfv\'en waves via mode conversion ($a=1/6$, $b=7/6$) and anomalous conduction damping ($a=-0.5$, $b=-0.5$),  the thermal mode is always unstable in the case of suppressed conduction, and there are combinations of the coronal densities and temperatures for which the thermal mode can be either stable or unstable in the case of finite conduction (see Fig.~\ref{fig:ab_regions}). Likewise, the models $a=-1$, $b=0$; $a=-0.5$, $b=-0.5$; and $a=0$, $b=0$ always cause damping of the acoustic mode in the regime of finite conductivity, while the other two models lead to the acoustic overstability for certain densities and temperatures.\\


\section{Comparison of the thermal misbalance time scales to observed wave periods}
\label{sec:estim}
\begin{figure}
	\centering
	\includegraphics[width=\linewidth]{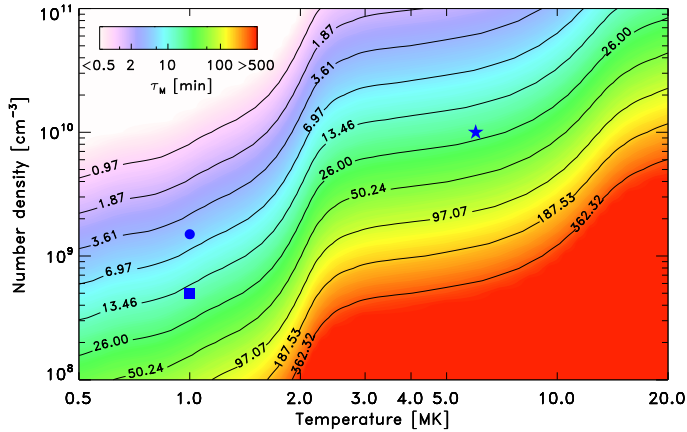}
	\caption{Characteristic thermal misbalance time $\tau_\mathrm{M}$ (\ref{eq:tau_m}) for typical coronal densities and temperatures and a fixed heating model $a=0.5$ and $b=-3.5$, for which both acoustic and thermal modes are stable over the entire intervals of plasma densities and temperatures considered (see Fig.~\ref{fig:ab_regions}). The colour scheme is adjusted so that the values of $\tau_\mathrm{M}$ from about $10$\,min to $100$\,min are shown in green. The blue symbols illustrate typical combinations of coronal plasma density and temperature in hot and dense loops in active regions (star), warm and less dense quiescent loops (circle), and polar plumes (box).
	}
	\label{fig:tau_rho_T}
\end{figure}

\begin{figure*}
	\centering
	\includegraphics[width=0.32\linewidth]{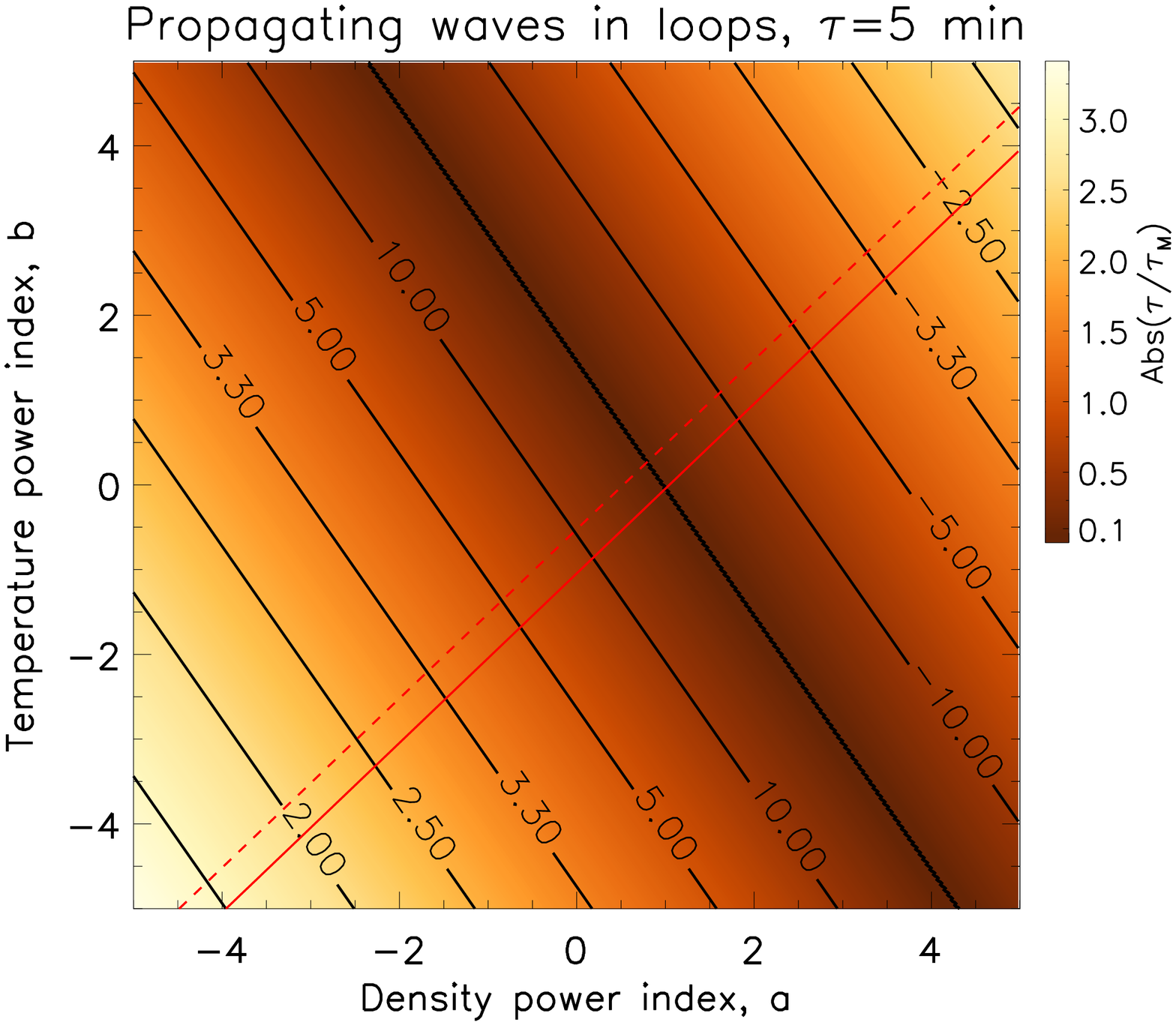}
	\includegraphics[width=0.32\linewidth]{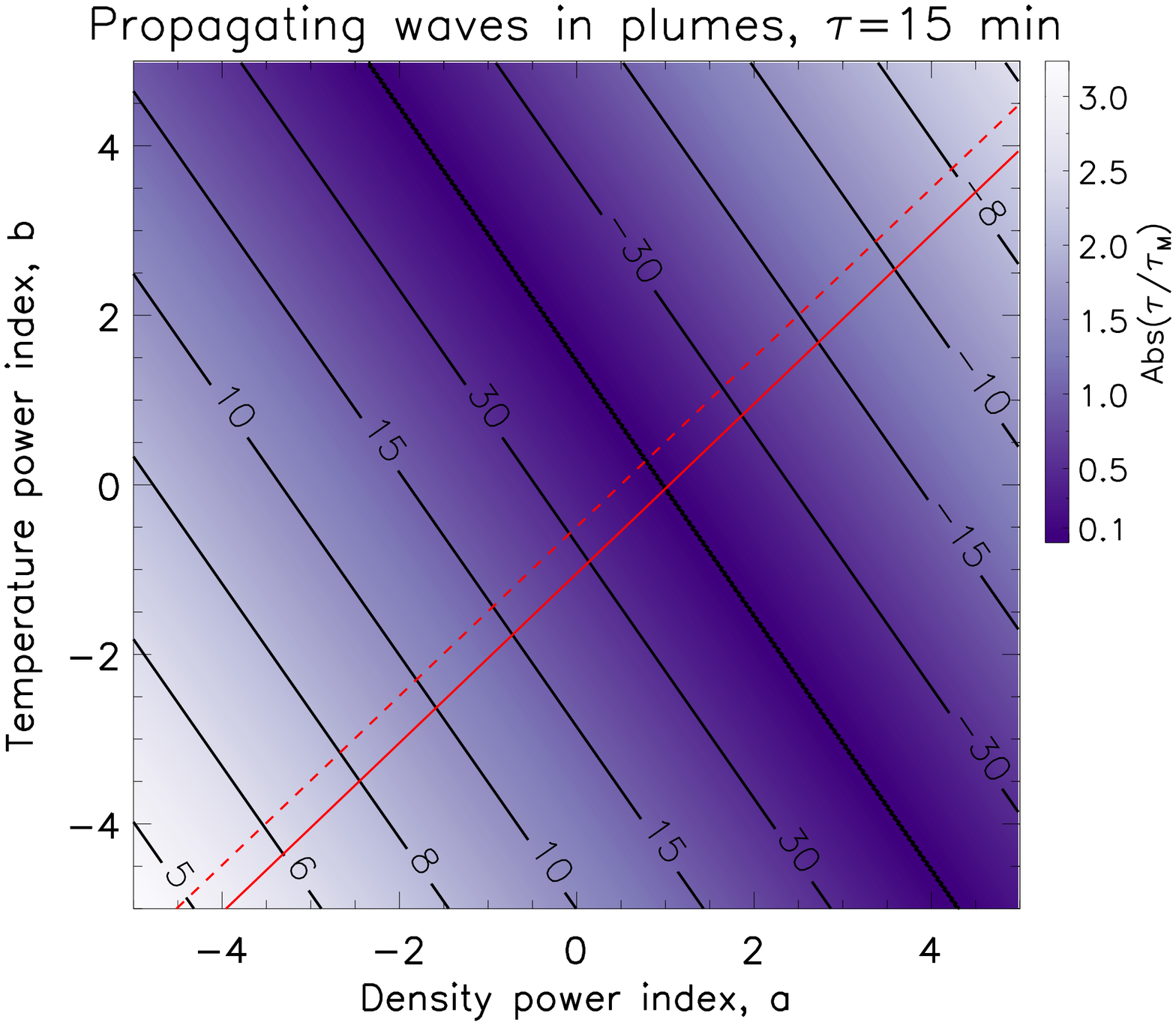}
	\includegraphics[width=0.32\linewidth]{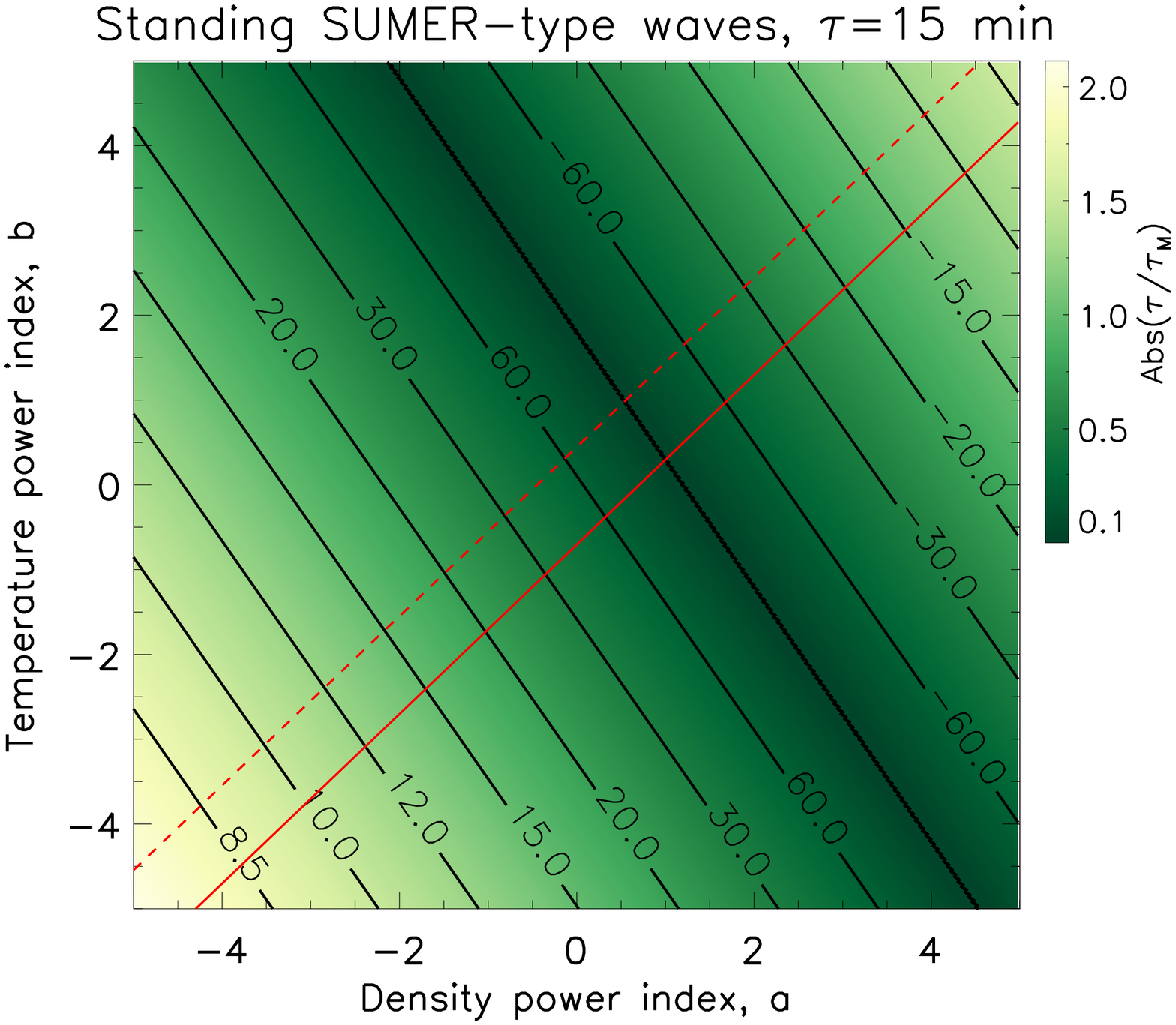}
	\caption{Estimations of $\tau_\mathrm{M}$ (the black contours, in minutes) for varying heating models (power indices $a$ and $b$) in warm and less dense quiescent loops (left), polar plumes (middle), and hot and dense loops (right).
		For specific combinations of the plasma density and temperature for these cases see Sec.~\ref{sec:estim}.
		The colour schemes in all panels show the ratio of $\tau_\mathrm{M}$ to typical periods $\tau$ of slow waves observed in those structures. The red solid lines show the values of $a$ and $b$, above which the thermal mode becomes unstable. The red dashed lines indicate the intervals where the thermal mode instability growth time is up to 5 times longer than the corresponding acoustic period $\tau$.
	}
	\label{fig:tau_ab}
\end{figure*}

We can combine the time scales $\tau_1$ and $\tau_2$ in Eq.~(\ref{eq:instab_cond_ac}) into a single misbalance time
\begin{equation}\label{eq:tau_m}
\tau_\mathrm{M}=\frac{\tau_1\tau_2}{\tau_1-\tau_2},
\end{equation}
and estimate it for different combinations of plasma density and temperature (Fig.~\ref{fig:tau_rho_T}) and for varying parameters of the heating function (Fig.~\ref{fig:tau_ab}).

Figure~\ref{fig:tau_rho_T} shows variation of the misbalance time $\tau_\mathrm{M}$ (\ref{eq:tau_m}) with plasma density and temperature for a fixed heating model $a=0.5$ and $b=-3.5$. In this illustrative example both the acoustic and thermal modes are found to damp for all considered plasma densities and temperatures (see Fig.~\ref{fig:ab_regions}).
As shown by Fig.~\ref{fig:tau_rho_T}, the time $\tau_\mathrm{M}$ varies from a few minutes to several tens of minutes for typical combinations of the plasma density and temperature. In particular, the value of $\tau_\mathrm{M}$ is similar, about several minutes, both in hot and dense plasmas, e.g., with $T_0\simeq10$\,MK and $n_0\simeq10^{10}$\,cm$^{-3}$, and in a cooler and less dense plasma, e.g., with $T_0\simeq1$\,MK and $n_0\simeq10^{9}$\,cm$^{-3}$. These values of $\tau_\mathrm{M}$ are similar to the slow wave periods detected in typical coronal plasma structures (see Table~\ref{tab:params}).
In contrast, in very dense and cool or very hot and rarified plasmas, $\tau_\mathrm{M}$ is either very short ($<0.5$\,min) or long ($>500$\,min).

For the analysis of the dependence of $\tau_\mathrm{M}$ on the heating function power indices $a$ and $b$, we choose three typical combinations of the plasma density and temperature, corresponding to typical long-lived, i.e. {apparently stable to instabilities}, plasma structures supporting slow waves: hot and dense loops in active regions ($T_0=6$\,MK, $n_0=10^{10}$\,cm$^{-3}$), warm and less dense quiescent loops ($T_0=1$\,MK, $n_0=1.5\times10^{9}$\,cm$^{-3}$), and plumes ($T_0=1$\,MK, $n_0=0.5\times10^{9}$\,cm$^{-3}$).
Slow waves observed in those structures have typical periods of several minutes {and damping times comparable to the oscillation periods} (see also Table~\ref{tab:params}).

Values of the misbalance time $\tau_\mathrm{M}$ and its ratio to the typical wave periods $\tau$ detected in corresponding plasma structures are shown in Fig.~\ref{fig:tau_ab}. According to Eqs.~(\ref{eq:instab_cond_ac}) and (\ref{eq:instab_cond_ac_ab}), the acoustic mode damps for $\tau_\mathrm{M}>0$ and grows for $\tau_\mathrm{M}<0$, that correspond to two distinct regions in the parametric plane ($a$, $b$). We also use Eq.~(\ref{eq:instab_cond_th_ab}) for obtaining the values of $a$ and $b$ for which the heating/cooling misbalance leads to thermal mode instability. Moreover, we solve the full dispersion relation (\ref{eq:dr}) numerically for obtaining the intervals of $a$ and $b$ for which the growth time of the thermal-mode instability is long enough, e.g., 5 times longer than the acoustic period $\tau$, to have just a mild effect on the dynamics of the acoustic mode. The latter could be referred to as a soft stability of thermal mode. Accounting for all these conditions in Fig.~\ref{fig:tau_ab}, we obtain that for a reasonable choice of the heating model, i.e. for $-1\lesssim a\lesssim0.5$ and $-1.5\lesssim b\lesssim-0.5$ for propagating slow waves in loops and plumes and for $-2\lesssim a\lesssim 0$ and $-1.5\lesssim b\lesssim-0.5$ for standing SUMER-type oscillations, the thermal mode is in the stable or soft-stable regime, and the values of the misbalance time $\tau_\mathrm{M}$ are either equal to or up to a factor of two longer than the acoustic wave period $\tau$, indicating a strong effect of the heating/cooling misbalance on the wave dynamics in these regimes.
{For example as shown by Fig.~\ref{fig:tau_ab}, the heating model by Alfv\'en waves via anomalous conduction damping ($a=-0.5$ and $b=-0.5$) from \citet{1993ApJ...415..335I} appears to cause strong damping (with the damping time $\approx \tau_\mathrm{M}$ being about the wave oscillation period) of standing SUMER-type oscillations in flaring loops, while the oscillating loop itself remains soft-stable to the thermal instability.}
We also note that these estimations of $a$ and $b$ are fully based on the effect of the heating/cooling misbalance, and can be softened by finite thermal conductivity and other dissipative processes.\\

\section{Discussion and conclusions} \label{sec:discuss}
The coronal heating function can be constrained seismologically using observations of damped slow magnetoacoustic waves in long-lived coronal plasma structures. The developed approach is based on the effect of the heating/cooling misbalance, that occurs due to the violation of the coronal thermal equilibrium between heating and cooling processes by {compressive} (e.g., slow) magnetoacoustic waves, causing a back reaction on the waves, which can either attenuate or amplify the wave. The latter leads to acoustic overstability or thermal instability, which are incompatible with the observed lifetimes of typical thermodynamically stable coronal structures \citep[cf. the phenomenon of catastrophic cooling due to thermal instability, e.g.][]{1988SoPh..117...51D,2010ApJ...716..154A}. 

Parametrising the coronal heating function as ${\cal H}(\rho,T)\propto \rho^aT^b$, we derived and evaluated conditions for the slow magnetoacoustic (acoustic) overstability and thermal instability {in the infinite magnetic field approximation, in terms of the power-indices $a$ and $b$ (Eqs.~(\ref{eq:instab_cond_ac_ab})--(\ref{eq:instab_cond_th_ab}))}. The heating models outlined by a triangle with approximate vertices $(-2.5,-5)$, $(1,-1.5)$, and $(3.5,-5)$ in the parametric plane ($a$, $b$) allow for a stable corona (see Fig.~\ref{fig:ab_regions}). In particular, the five heating models from \citet{1993ApJ...415..335I} are all shown to be potentially unstable to the thermal mode in the regime of suppressed thermal conduction, and there are certain combinations of coronal plasma densities and temperatures for which some of those heating models can lead to acoustic and thermal stability of coronal structures with finite thermal conduction.

We showed that characteristic times of the thermal misbalance, $\tau_{1,2}$ (\ref{eq:dr}), and their specific combination $\tau_\mathrm{M}$ (\ref{eq:tau_m}), determined by the dependences of the cooling and heating rates on the plasma parameters perturbed by the wave, are comparable to the slow wave oscillation periods and damping times (about several minutes) {typical for the corona}, in a broad range of typical coronal plasma temperatures and densities (see Fig.~\ref{fig:tau_rho_T}). This demonstrates a broad importance and applicability of this effect for probing heating functions in various specific coronal structures. We also note that those misbalance time scales are independent of the amplitudes of the heating ${\cal H}_0$ and cooling ${\cal L}_0$ processes in the linear regime and describe how quickly the plasma perturbed by a compressive wave returns to the thermal equilibrium or becomes thermodynamically unstable (cf. the radiative cooling time $\tau_\mathrm{rad}$ (\ref{eq:tau_rad}), which depends on ${\cal L}_0$ and shows how quickly the hot plasma cools down by radiation if for some reason the heating process switches off).

We compared the thermal misbalance time scale $\tau_\mathrm{M}$ (\ref{eq:tau_m}) with observed oscillation periods and damping times of slow waves in quiescent and flaring loops, and plumes in coronal holes. Requesting the thermal mode to be either stable or at least to grow at a relatively low rate (soft-stable), we obtained that the heating models with $-1\lesssim a\lesssim0.5$ and $-1.5\lesssim b\lesssim-0.5$ (for propagating slow waves in quiescent loops and polar plumes) and $-2\lesssim a\lesssim 0$ and $-1.5\lesssim b\lesssim-0.5$ (for standing SUMER-type oscillations in flaring loops) can cause a strong damping consistent with observations, with the thermal misbalance time scale $\tau_\mathrm{M}$ being about observed slow wave periods (see Fig.~\ref{fig:tau_ab}).
{The empirically revealed negative values of the parameter $b$ imply that coronal plasma with colder temperatures (at constant density) would require a greater heating rate to be thermally and acoustically stable.}
{The observation of slow magnetoacoustic oscillations in coronal plasma structures with different combinations of the density and temperature would help to improve the estimation of the heating function.}

Thus, we demonstrated that the effect of the thermodynamic activity of the corona is strong and crucial for modelling and interpreting observed slow magnetoacoustic waves and has {promising seismological implications}. It opens up a new perspective for the use of observations of slow {waves for} inferring thermodynamic parameters of unknown coronal heating mechanisms. Our study is based on the linear analysis in a plasma uniform along the magnetic field.
{In this regime, the effect of the density non-uniformities generated by the wave itself is neglected as a higher-order effect. Also,} the effects of the nonlinear stabilisation of linearly unstable structures, and non-linear instabilities of linearly stable structures, as well as the plasma non-uniformity could be considered as the next step.

\begin{acknowledgements}
The work was supported by the STFC consolidated grants ST/P000320/1 and ST/T000252/1. V.M.N. was supported by the Russian Foundation for Basic Research Grant No.~18-29-21016. D.Y.K. was supported by the budgetary funding of Basic Research program No. II.16.
CHIANTI is a collaborative project involving George Mason University, the University of Michigan (USA) and the University of Cambridge (UK).
\end{acknowledgements}

\bibliographystyle{aa} 

\end{document}